\documentclass[rmp, sort&compress,numbers]{revtex4}
\usepackage{graphicx}
\usepackage{rotate}
\usepackage{epsfig}

\usepackage{color}

\usepackage{amssymb,amsfonts,amsmath}

\begin{document}

\title{Activity driven modeling of time varying networks}

\author{Nicola Perra${}^{1,2}$, Bruno Gon\c calves${}^{1}$, Romualdo Pastor-Satorras${}^{3}$, Alessandro Vespignani${}^{1,4,5}$}
\affiliation{${}^{1}$Department of Physics, College of Computer and Information Sciences, Department of Health Sciences, Northeastern University, Boston MA 02115 USA\\
    ${}^{2}$Linkalab, Cagliari, Italy\\
    ${}^{3}$Departament de F\'{\i}sica i Enginyeria Nuclear, Universitat Polit\`ecnica de Catalunya,
    Campus Nord B4, 08034 Barcelona, Spain\\
    ${}^{4}$Institute for Scientific Interchange
    Foundation, Turin 10133, Italy\\
    ${}^{5}$Institute for
    Quantitative Social Sciences, Harvard University, Cambridge, MA,
    02138\\} 

\begin{abstract}
Network modeling plays a critical role in identifying statistical regularities and structural principles common to many systems. The large majority of recent modeling approaches are connectivity driven. The structural patterns of the network are at the basis of the mechanisms ruling the network formation. Connectivity driven models necessarily provide a time-aggregated representation that may fail to describe the instantaneous and fluctuating dynamics of many networks.  We address this challenge by defining the activity potential, a time invariant function characterizing the agents' interactions and constructing an activity driven model capable of encoding the instantaneous time description of the network dynamics. The model provides an explanation of structural features such as the presence of hubs, which simply originate from the heterogeneous activity of agents. Within this framework, highly dynamical networks can be described analytically, allowing a quantitative discussion of the biases induced by the time-aggregated representations in the analysis of dynamical processes.
\end{abstract}

\maketitle

Network
modeling~\cite{newman10-1,barrat08-1,albert02,boccaletti06-1,bollobas98-1,vespignani12-1}
has long drawn on the tradition of social network analysis and graph
theory, with models ranging from the Erd\"os-R\'enyi model to Logit
models, p*-models, and Markov random graphs \cite{erdos59,
  molloy95-1,holland81-1,franc86-1,wasserman96-1}. In the last decade,
the class of growing network models, exemplified by the preferential
attachment model, has been made widely popular by research in
statistical physics and computer science \cite{barabasi99-2,
  barabasi99-3, dorogovtsev00-1,mendes-book,
  fortunato06-1,boguna11-1}. All these models can be defined as
connectivity driven, as the network's topology is at the core of the
model's algorithmic definition. Connectivity-driven network models are
well-suited for capturing the essential features of systems such as
the Internet, where connections among nodes are long-lived
elements~\cite{psatorras04-1,albert99}. However, in many cases the
interactions among the elements of the system are rapidly changing and
are characterized by processes whose timing and duration are defined
on a very short time scale~\cite{holme11-1,ghoshal05-1}. This limit has been investigated in the case of adaptive systems whose structure evolve being coupled to the process taking place on top of them~\cite{volz09-1,centola07-1,jolad11-1,schwartz10-1,schwartz10-2}. Instead, the understating of this limit in time varying networks in which the structure evolves independently of the process is still limited and unexplored.  In these activity-driven networks, models
intended to capture the process of accumulating connections over time
and the resulting degree distribution (i.e. the probability that a
node has $k$ connections to other nodes) and other topological
properties merely represent a time-integrated perspective of the
system. Furthermore, the analysis of dynamical processes in evolving
networks is generally performed in the presence of a time-scale
separation between the network evolution and the dynamical process
unfolding on its structure. In one limit we can consider the network
as quenched in its connectivity pattern, thus evolving on a time scale
that is much longer that the dynamical process itself. In the other
limiting case, the network is evolving at a time scale much shorter
than the dynamical process thus effectively disappearing from the
definition of the interaction among individuals that is conveniently
replaced by effective random couplings. While the time scale
separation is extremely convenient for the numerical and analytical
tractability of the models, networks generally evolve on a time-scale
that might be comparable to the one of the dynamical
process~\cite{butts09-1,butts08-1,panisson11-1,moody02-1,morris97-1,basu}. An
accurate modelization of the dynamics of activity-driven networks
calls for the definition of interaction processes based on the actual
measurement of the activity of the agents forming the system, a task
now made possible by the availability of time-resolved, high-quality
data on the instantaneous activity of millions of agents in a wide
variety of networks \cite{gonzalez08-1,
  onnela06-1,lazer09-1,vespignani09-1,brockmann06-1}.
  
\section{Results} 

Here we present the analysis of three large-scale, time-resolved
network datasets and define for each node a measurable quantity, the
activity potential, characterizing its interaction pattern within the
network. This measure is defined as the number of interactions
performed, in a given time window, by each node divided by the total
number of interactions made by all the nodes in the same time
window. We find that the system level dynamics of the network can be
encoded by the activity potential distribution function from which it
is possible to derive the appropriate interaction rate among nodes. On
the basis of the empirically measured activity potential distribution
we propose a process model for the generation of random dynamic
networks.  The activity potential function defines the network
structure in time and traces back the origin of hubs to the
heterogenous activity of the network elements. The model allows to
write dynamical equations coupling the network dynamics and the
dynamical processes unfolding on its structure without relying on any
time-scale separation approximation. We analyze a simple spreading
process and provide the explicit analytical expression for the biases
introduced by the time-aggregated representation of the network when
studying dynamical processes occurring on a time scale comparable to
that of the network evolution.  Interestingly the network model
presented here is amenable to the introduction of many features in the
nodes' dynamic such as the the persistency of specific interactions or
assortative/disassortative correlations, thus defining a general basic
modeling framework for rapidly evolving networks.\\

\begin{figure*}
\begin{center}
\includegraphics[width=0.9\textwidth,angle=0]{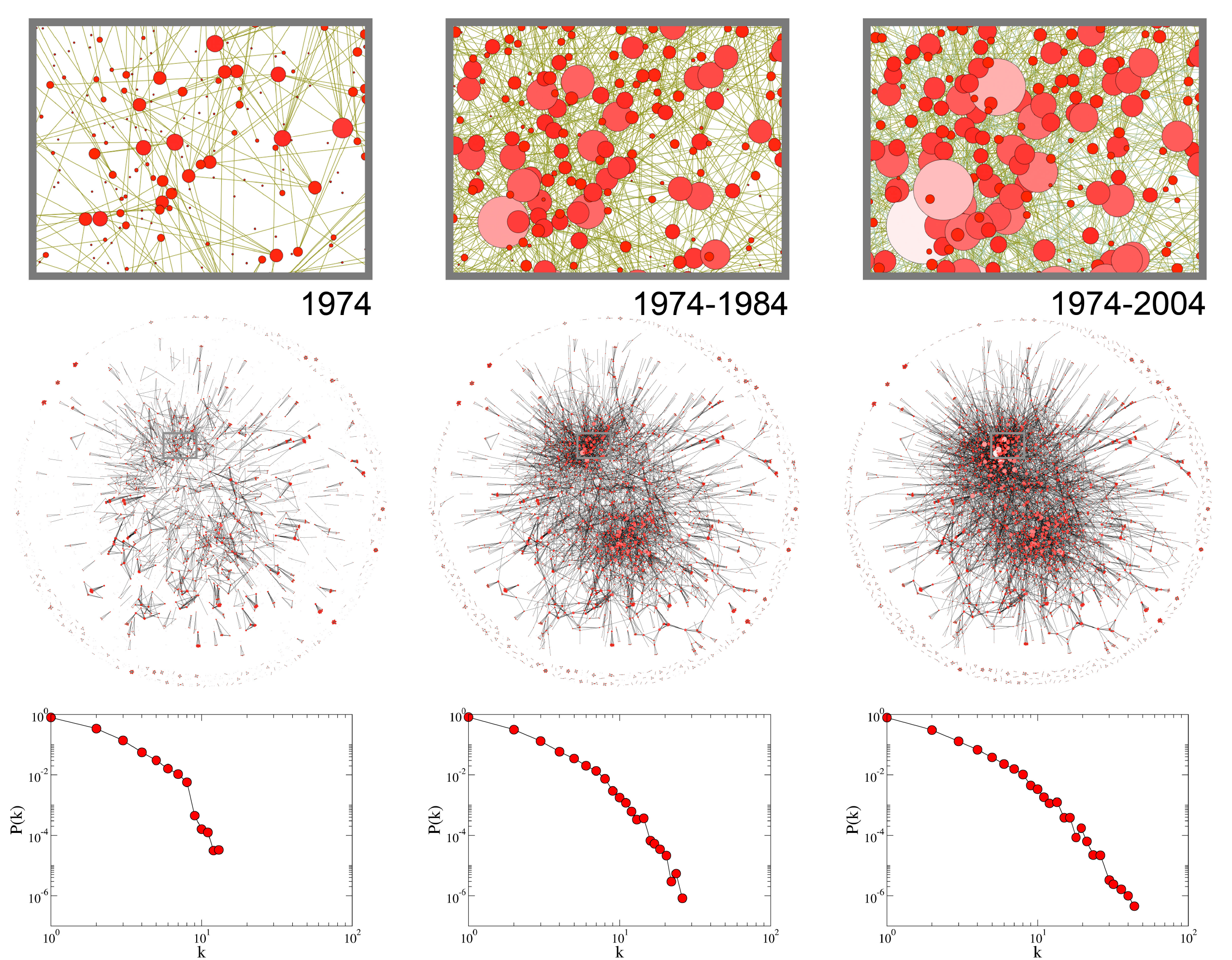}
\end{center}
\caption{ Network visualization and degree distribution of the PRL
  dataset considering three different aggregated views. In particular,
  in the first two rows we focus on the set of authors who wrote at
  least one paper in the period between $1960$ and $1974$. For this
  subset of $5,162$ active authors we construct three different
  networks, graphically represented in the central row of the
  figure. The upper row represents a blown up perspective of a
  particular network region. In the left column we show the network of
  $1974$, defined by the active nodes in the given time frame. The
  central column shows the network obtained by integrating over $10$
  years, from $1974$ to $1984$.  In the right column we show the
  network obtained by integrating over $30$ years, from $1974$ to$
  2004$.  The first network is highly fragmented as is obvious from
  the visualization. When larger windows are integrated the density of
  the network increases and heterogeneous connectivity patterns start
  to emerge. Clearly, as indicated by the degree distributions, that
  consider the complete set of authors (not just those used for the
  sake of visualization in the first two rows), the time scale used
  to construct the network affects its topological structure. In each
  visualization the size and color of the nodes is proportional to
  their degree.}
\label{fig1}
\end{figure*}

\subsection{The activity potential.}
We consider three datasets corresponding to networks in which we can
measure the individual agents' activity: Collaborations in the journal
"Physical Review Letters" (PRL) published by the American Physical
Society~\cite{aps10-1}, messages exchanged over the Twitter microblogging
network, and the activity of actors in movies and TV series as
recorded in the Internet Movie Database (IMDb)~\cite{IMBD10-1}.  In
the first dataset the network representation considers undirected
links connecting two PRL authors if they have collaborated in writing
one article. In the second system each node is a Twitter user and an
undirected link is drawn if at least one message has been exchanged
between two users. Finally, the actor network is obtained by drawing
an undirected link between any two actors who have participated in the
same movie or TV series.

Simple evidence for the role of agents' activity in network analysis
and modeling can be readily observed in the case of the collaboration
network of scientific authors\cite{newman01-4}. The number of
collaborations of any author depends on the time window through which
we observe the system. In Fig.\ref{fig1} we show the networks obtained by
time-aggregated co-authorships over $1$, $10$, and $30$ years for the
subset of authors in the PRL dataset who were active in the considered
time period. Clearly, the time scale used to construct the network
defines a non-stationary connectivity pattern and explicitly affects
the network structure and its degree distribution. Similar results are
found for the other two datasets as shown in the Supplementary
Information.

\begin{figure*}
\includegraphics[width=0.55\columnwidth,angle=0]{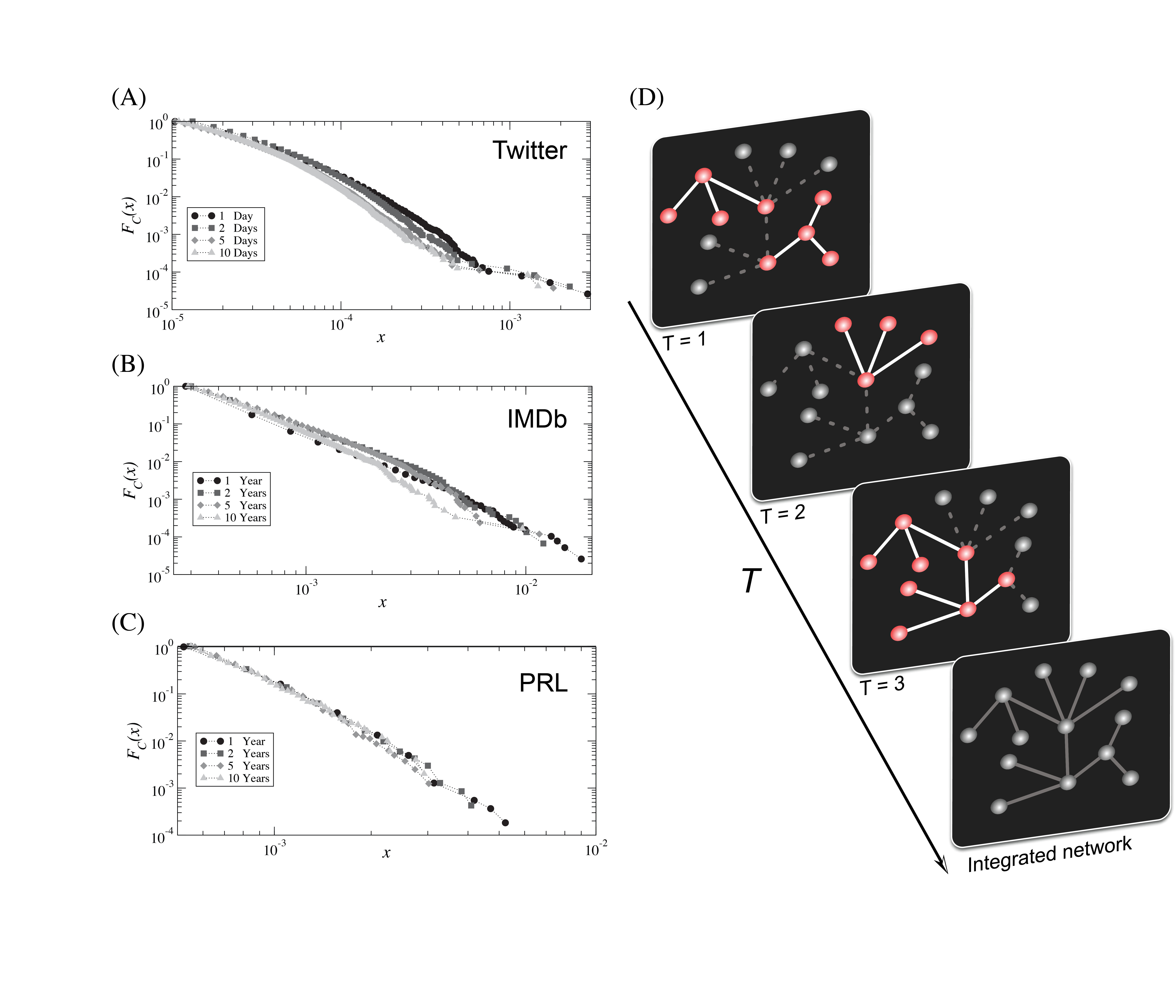}
\caption{Cumulative distribution of the activity potential, $F_C(x)$, empirically measured by using  four different time windows and a schematic representation of the proposed network model. In particular, in panel (A) we show the cumulative distributions of the observables $x$ for Twitter, in panel (B) for IMDb, and in panel (C) for PRL. In panel (D) we show a schematic representation of the model. Considering just $13$ nodes and $m=3$, we plot a visualization of the resulting networks for $3$ different time steps. The red nodes represent the firing/active nodes. The final visualization represents the network after integration over all time steps.}
\label{fig2}
\end{figure*}

In the three datasets considered, we characterize the individual
activity of every agent: papers written, messages exchanged, or movie
appearances, respectively.  For each dataset we measure the individual
activity of each agent and define the \emph{activity potential} $x_i$
of the agent $i$ as the number of interactions that he/she performs in
a characteristic time window of given length $\Delta t$, divided by
the total number of interactions made by all agents during the same
time window. The activity potential $x_i$ thus estimates the
probability that the agent $i$ was involved in any given interaction
in the system, and the probability distribution $F(x)$ that a randomly
chosen agent $i$ has activity potential $x$ statistically defines the
interaction dynamics of the system. In Fig.~\ref{fig2} we show the
cumulative distribution $F_c(x)$ evaluated for the three datasets. In
all cases we find that, contrary to the degree distribution and other
structural characteristics of the networks, the distribution $F_c(x)$
is virtually independent of the time scale over which the activity
potential is measured. Additionally, we find that the distribution
$F_c(x)$ is skewed and fairly broadly distributed. This is hardly
surprising as in many cases measurements of human activity have
confirmed the presence
of wide variability across individuals~\cite{barabasi05-1,marton12}.\\

\subsection{Activity driven network model.}  Our empirical analysis naturally
leads to the definition of a simple model that uses the activity
distribution to drive the formation of a dynamic network. We consider
$N$ nodes (agents) and assign to each node $i$ an activity/firing rate
$a_i = \eta x_i$, defined as the probability per unit time to create
new contacts or interactions with other individuals, where $\eta$ is a
rescaling factor defined such that the average number of active nodes
per unit time in the system is $\eta \langle x \rangle N$. The
activity rates are defined such that the numbers $x_i$ are bounded in
the interval $\epsilon \le x_i \le1$, and are assigned according to a
given probability distribution $F(x)$ that may be chosen arbitrarily
or given by empirical data. We impose a lower cut-off $\epsilon$ on
$x$ in order to avoid possible divergences of $F(x)$ close to the
origin. We assume a simple generative process according to the
following rules (see Fig.\ref{fig2}-D):
\begin{itemize}
\item At each discrete time step $t$ the network $G_t$ starts with $N$
  disconnected vertices;
\item With probability $a_i \Delta t$ each vertex $i$ becomes active
  and generates $m$ links that are connected to $m$ other randomly
  selected vertices. Non-active nodes can still receive connections
  from other active vertices;
\item At the next time step $t + \Delta t$, all the edges in the
  network $G_t$ are deleted. From this definition it follows that all
  interactions have a constant duration $\tau_i = \Delta t$.
\end{itemize}
The above model is random and Markovian in the sense that agents do
not have memory of the previous time steps. The full dynamics of the
network and its ensuing structure is thus completely encoded in the
activity potential distribution $F(x)$.

\begin{figure*}
\includegraphics[width=0.9\columnwidth,angle=0]{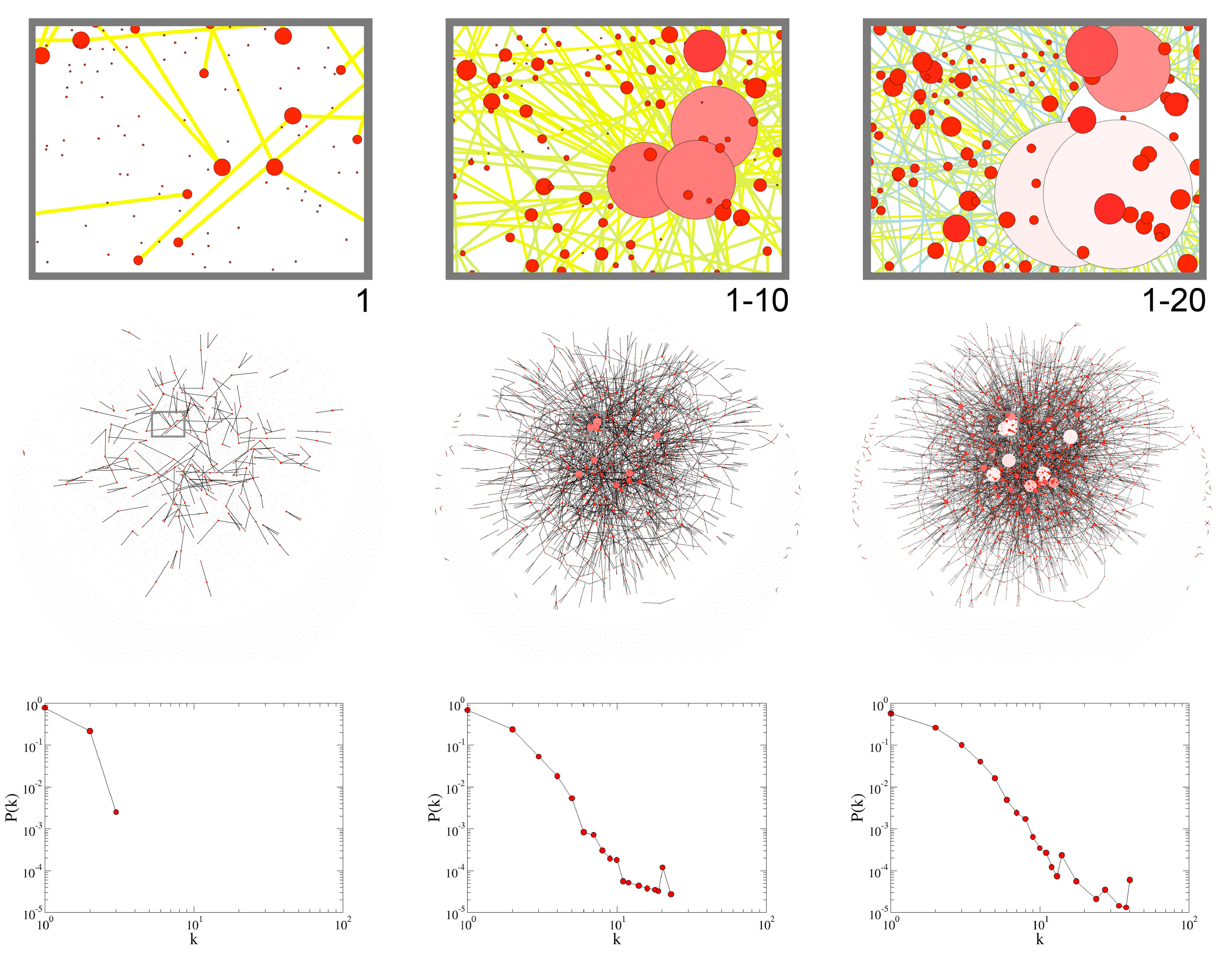}
\caption{Visualization and degree distributions of the proposed network model considering different aggregated views. We fix $N=5000$, $m=2$, $\eta=10$, $F(x) \propto x^{-\gamma}$ with $\gamma =2.8$, $\epsilon \le x \le 1$ with $\epsilon =10^{-3}$. We plot the network obtained after one time step in the first column, the network obtained after integrating over $10$ iterations in the second column, and the network obtained after integrating over $20$ iterations in the last column. Interestingly, even though the model is random and markovian by construction, we observe a behavior qualitatively similar to the case of PRL: the single time window yields a sparse and poorly connected network with a trivial degree distribution. When larger time scales are considered, heterogeneous connectivity patterns start to emerge as seen by the corresponding degree distributions. In each visualization the size and color of the nodes is proportional to their degree.}
\label{fig3}
\end{figure*}

In Fig.~\ref{fig3} we report the results of numerical simulations of a network
with $N =5000$, $m =2$, $\eta =10$, and $F(x) \propto x^{-\gamma} $,
with $\gamma =2.8$ and $\epsilon =10^{-3}$. The model recovers the
same qualitative behavior observed in Fig.~\ref{fig1}. At each time step the
network is a simple random graph with low average connectivity. The
accumulation of connections that we observe by measuring activity
on increasingly larger time slices $T$ generates a skewed $P_T(k)$
degree distribution with a broad variability. The presence of
heterogeneities and hubs (nodes with a large number of connections) is
due to the wide variation of activity rates in the system and the
associated highly active agents. However, it is worth remarking that
hub formation has a different interpretation than in growing network
prescriptions, such as preferential attachment. In those cases hubs
are created by a positional advantage in degree space leading to the
passive attraction of more and more connections. In our model, the
creation of hubs results from the presence of nodes with high activity
rate, which are more willing to repeatedly engage in interactions.

The model allows for a simple analytical treatment. We define the
integrated network $G_T=\bigcup_{t=0}^{t=T} {G}_t$ as the union of all
the networks obtained in each previous time step. The instantaneous
network generated at each time $t$ will be composed of a set of
slightly interconnected nodes corresponding to the agents that were
active at that particular time, plus those who received connections
from active agents. Each active node will create $m$ links and the
total edges per unit time are $ E_t= m N \eta \langle x \rangle$
yielding the average degree per unit time Ð the contact rate Ð of the
network 
\begin{equation}
 \langle k \rangle_t=\frac{2E_t}{N}=2m\eta \langle x
\rangle.  
\end{equation} 

\begin{figure*}
\includegraphics[width=0.55\columnwidth,angle=0]{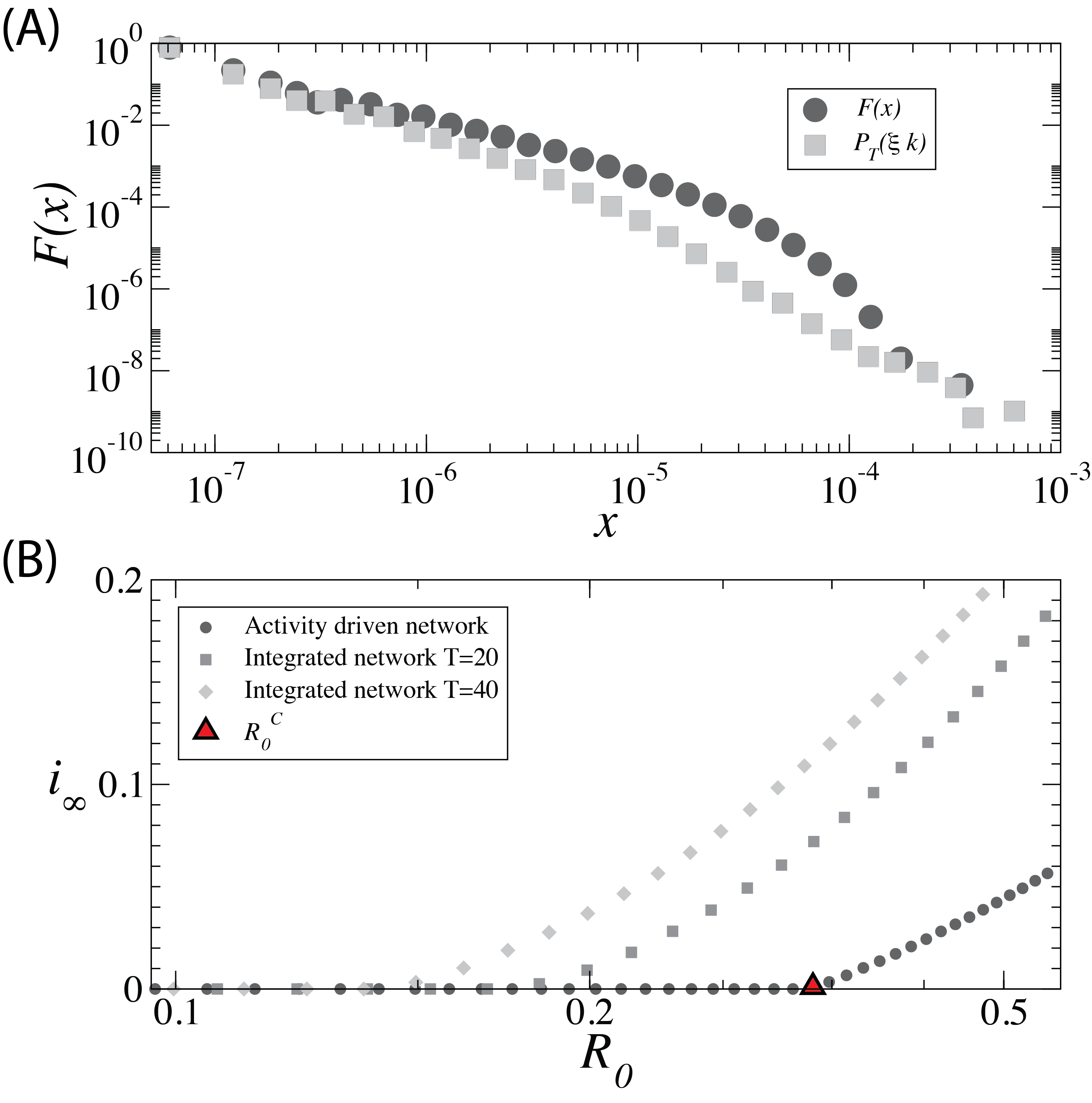}
\caption{In panel (A) we consider the entire Twitter dataset and show the distribution of activity potential $F(x)$ and the asymptotic degree distribution of the corresponding network, $P_T[\xi k]$, with $\xi=1/(T\eta m)$, rescaled according to the analytical result. In panel (B) we show the density of infected nodes, $I_\infty$, in the stationary state, obtained from numerical simulations of the SIS model on a network generated according to the proposed model and two other networks resulting from an integration of the model over $20$ and $40$ time steps, respectively. We set $N=10^6$, $m=5$, $\eta=10$, $F(x) \propto x^{-\gamma}$ with $\gamma =2.1$ and $\epsilon \le x \le 1$ with $\epsilon =10^{-3}$. Each point represents an average over $10^2$ independent simulations. The red triangle marks the critical reproductive number $R_o^c$ as predicted by Eq.~\ref{thre}.}
\label{fig4}
\end{figure*}

The instantaneous network will be composed by a set of
stars, the vertices that were active at that time step, with degree
larger than or equal to $m$, plus some vertices with low degree. The
corresponding integrated network, on the other hand, will generally
not be sparse, being the union of all the instantaneous networks at
previous times (see Fig.~\ref{fig3}). In fact, for large time $T$ and network
size $N$, when the degree in the integrated network can be
approximated by a continuous variable, we can show (see Supplementary
Information) that agent $i$ will have at time $T$ a degree in the
integrated network given by $k_i(T)=N \left (1- e^{-Tm\eta x_i/N}
\right)$. It can then easily be shown that the degree distribution
$P_T(k)$ of the integrated network at time $T$ takes the form:

 \begin{equation}
P_T(k)\sim F\left[ \frac{k}{T m \eta}\right], 
\end{equation}

where we have considered the limit of small $k/N$ and $k/T$
(i.e. large network size and times). The noticeable result here is the
relation between the degree distribution of the integrated network and
the distribution of individual activity, which, from the previous
equation, share the same functional form. This relation is
approximately recovered in the empirical data, where the activity
potential distribution is in reasonable agreement with the
appropriately rescaled asymptotic degree distribution of the
corresponding network (see Fig.~\ref{fig4}-A). As expected,
differences between the two distributions are present, due to features
of the real network dynamics that our random model does not capture:
links might have memory (already explored connections are more likely
to happen again), social relations have a lifetime distribution
(persistence) and multiple connections and weighted links may be
relevant. Neither of these effects is considered in the model. We
report some statistical analysis of those features in the
Supplementary Information as further ingredients to be considered in
future extensions of the model.

\subsection{Dynamical processes in activity driven networks.}  Recent
research has highlighted the key role of interaction dynamics as
opposed to static studies. For example, an individual who appears to
be central by traditional network metrics may in fact be the last to
be infected because of the timing of his/her
interactions~\cite{moody02-1,isella11-1}. Analogously the concurrency
of sexual partners can dramatically accelerate the spread of STDs
\cite{morris97-1}. Despite its simplicity, our model makes it
analytically explicit that the actors' activity time scale plays a
major role in the understanding of processes unfolding on dynamical
networks.  Let us consider the susceptible-infected-susceptible (SIS)
epidemic compartmental
model~\cite{kermac27-1,newman10-1,barrat08-1,keeling08-1}. In this
model, infected individuals can propagate the disease to healthy
neighbors with probability $\lambda$, while infected individuals
recover with rate $\mu$ and become susceptible again. In an homogenous
population the behavior of the epidemics is controlled by the
reproductive number $R_0 = \beta/\mu $, where $\beta = \lambda \langle
k \rangle$ is the per capita spreading rate that takes into account
the rate of contacts of each individual. The reproductive number
identifies the average number of secondary cases generated by a
primary case in an entirely susceptible population and defines the
epidemic threshold such that only if $R_0>1$ can epidemics reach an
endemic state and spread into a closed population.  In the past few
years the inclusion of complex connectivity networks and mobility
schemes into the substrate of spreading processes Ð contagion,
diffusion, transfer, etc. has highlighted new and interesting
results~\cite{lloyd01-1,balcan09-1,wang03-1,chakrabarti08-1,castellano10-1}.
Several results states that the epidemic threshold depends on the
topological properties of the networks. In particular, for networks
characterized by a fix, quenched topology the threshold is given by
the principal eigenvalue of the adjacency
matrix~\cite{wang03-1,chakrabarti08-1}. Instead, for annealed network,
characterized by a topology defined just on average because the
connectivity patterns has a dynamic extremely fast with respect to the
dynamical process, heterogeneous mean-field
approaches~\cite{barrat08-1,vespignani12-1} predict an epidemic threshold that is
inversely proportional to the second moment of the network's degree
distribution: $\beta/ \mu > \langle k \rangle ^2 / \langle
k^2\rangle$. However, these results do not apply to the case in which
the time variation of the connectivity pattern is occurring on the
same time scale of the dynamical process. Our model presents simple
evidence of this problem, as a disease with a small value of
$\mu^{-1}$ (the infectious period characteristic time) will have time
to explore the fully-integrated network, but will not spread on the
dynamic instantaneous networks whose union defines the integrated
one~\cite{moody02-1,morris97-1,morris07-1,isella11-1}. In
Fig.~\ref{fig4}-B we plot the results of numerical simulations of the
SIS model on a network generated according to our model and on two
time-aggregated network instances. We observe that the two aggregated
networks lead to misleading results in both the threshold and the
epidemic magnitude as a function of $\beta/ \mu$. Even if the epidemic
threshold discounts the different average degree of the networks in
the factor $\beta =\lambda \langle k \rangle$, the two aggregated
instances consider all edges as always available to carry the
contagion process, disregarding the fact that the edges may be active
or not according to a specific time sequence defined by the agents'
activity.

The above finding can be more precisely quantified by calculating
analytically the epidemic threshold in activity driven networks
without relying on any time aggregated view of the network
connectivity. By working with activity rate we can derive epidemic
evolution equation in which the spreading process and the network
dynamics are coupled together.  Let us assume a distribution of
activity potential $x$ of nodes given by a general distribution $F(x)$
as before. At a mean-field level, the epidemic process will be
characterized by the number of infected individuals in the class of
activity rate $a$, at time $t$, namely $I^{t}_{a}$.  The number of
infected individuals of class $a$ at time $t+\Delta t$ given by:
\begin{equation}
  I^{t+\Delta t}_{a}= -\mu \Delta t I^{t}_{a} +I^{t}_{a}  +\lambda m (N_{a}^{t}-I_{a}^{t})a\Delta t \int d a' \frac{I_{a'}^{t}}{N}+\lambda m (N_{a}^{t}-I_{a}^{t})\int d a' \frac{I_{a'}^{t}a' \Delta t}{N},
\label{pp1}
\end{equation}
where $N_a$ is the total number of individuals with activity $a$. In
Eq.~(\ref{pp1}), the third term on the right side takes into account
the probability that a susceptible of class $a$ is active and acquires
the infection getting a connection from any other infected individual
(summing over all different classes), while the last term takes into
account the probability that a susceptible, independently of his
activity, gets a connection from any infected active individual.  The
above equation can be solved as shown in the material and methods
section, yielding the following epidemic threshold for the activity
driven model:
\begin{equation} 
\frac{\beta}{\mu} > \frac{2\langle a
  \rangle}{\langle a \rangle+\sqrt{\langle a^2 \rangle}}.
\label{thre}
\end{equation}

 This result considers the activity rate of each actor and
therefore takes into account the actual dynamics of interactions; the
above formula does not depend on the time-aggregated network
representation and provides the epidemic threshold as a function of
the interaction rate of the nodes. This allows to characterize the
spreading condition on the natural time scale of the combination of
the network and spreading process evolution.\\

\section{Discussion}
We have presented a model of dynamical networks that encodes the
connectivity pattern in a single function, the activity potential
distribution, that can be empirically measured in real world networks
for which longitudinal data are available. This function allows the
definition of a simple dynamical process based on the nodes' activity
rate, providing a time dependent description of the network's
connectivity pattern. Despite its simplicity, the model can be used to
solve analytically the co-evolution of the network and contagion
processes and characterize quantitatively the biases generated by
time-scale separation techniques. Furthermore the proposed model
appears to be suited as a testbed to discuss the effect of network
dynamics on other processes such as damage resilience, discovery and
data mining, collective behavior and synchronization. While we have
reduced the level of realism for the sake of parsimony of the
presented model, we are aware of the importance of analyzing other
features of actor activity such as concurrency, persistence and different weights associated with each connection. These features must necessarily be
added to the model in order to remove the limitations set by the
simple random network structures generated here and represent
interesting challenges for future work in this area.\\

\section{Methods}
\subsection{Datasets}

We considered three different dataset: the collaborations in the
journal "Physical Review Letters`` (PRL) published by the APS, the
message exchanged on Twitter and the activity of actors in movies and
TV series as recorded in the Internet Movie Database (IMDb). In
particular:
\\
\\
\emph{PRL dataset}. In this database the network representation
considers each author of a PRL article as a node. An undirected link
between two different authors is drawn if they collaborated in the
same article. We filter out all the articles with more than $10$
authors in order to focus our attention just on small collaborations
in which we can assume that the social components is relevant. We
consider the period between $1960$ and $2004$. In this time window we
registered $71,583$ active nodes and $261,553$ connections among them.
In this dataset is natural defining the activity rate, $a$, of each
author as the number of papers written in a specific time window
$\Delta t=1$ year.Authors with no collaborative papers in the total time span considered (isolates) are not included in the data set.
\\
\\
\emph{Twitter Dataset}. Having been granted temporary access to
Twitter's firehose we mined the stream for over $6$ months to identify
a large sample of active user accounts. Using the API, we then queried
for the complete history of $3$ million users, resulting in a total of
over $380$ million individual tweets covering almost $4$ years of user
activity on Twitter.  In this database the network representation
considers each users as a node. An undirected link between two
different users is drawn if they exchanged at least one message. We
focus our attention on $9$ months during $2008$. In this time window
we registered $531,788$ active nodes and $2,566,398$ connections among
them.  In this dataset we define the activity rate of each user as the
number of messages sent in a time window $\Delta t=1$ day.
\\
\\
\emph{IMDb Dataset}. In this database the network representation
considers each actor as a node. An undirected link between two
different actors is drawn if they collaborated in the same movie/TV
series. We focus on the period between $1950$ and $2010$. During this
time period we registered $1,273,631$ active nodes and $47,884,882$
connections between them.  A natural way to define the activity rate
in this dataset is to consider the number of movies acted by each
actor in a specific time window $\Delta t=$ 1 year.\\
\\
\subsection{Epidemic threshold}
In order to solve Eq.~(\ref{pp1}) we can consider the total number of
infectious nodes in the system 
\begin{equation}
\label{pp}
\int da I_{a}^{t+\Delta t}= I^{t+\Delta t}=I^{t}-\mu \Delta t I^{t} + \lambda m \langle a
\rangle \Delta t I^{t}+ \lambda m \theta^{t} \Delta t, 
\end{equation}
where $\theta^{t}=\int da' I^{t}_{a'}a' $ and we have dropped all
second order terms in the activity rate $a$ and in $I_a^{t}$.  We are not considering events in which two infected nodes
choose each other for connection and we are considering a linear approximation in $I_a^t$ since in the beginning of the epidemics the number of infectious individuals in each class is small. In order to obtain an
closed expression for $\theta$ we multiply both sides of
Eq.~(\ref{pp1}) by $a$ and integrate over all activity spectrum,
obtaining the equation
\begin{equation}
\theta^{t+\Delta t}=\theta^{t}-\mu \Delta t \theta^{t}+\lambda m \langle a^2 \rangle
I^{t} \Delta t+\lambda m \langle a \rangle\theta^{t}\Delta t. 
\label{othereq}
\end{equation}
In the continuous time limit we obtain the following closed system of equations 
\begin{eqnarray}
  \partial_{t} I &=&-\mu  I + \lambda m \langle a \rangle I + \lambda m \theta,\\
  \partial_{t} \theta &=& -\mu  \theta +\lambda m  \langle a^2\rangle
  I+\lambda m \langle a \rangle\theta ,
\end{eqnarray}
whose Jacobian matrix  has eigenvalues
\begin{equation}
\Lambda_{(1,2)} = \langle a \rangle\lambda m  - \mu  \pm \lambda m
\sqrt{\langle a^2\rangle }. 
\end{equation}
The epidemic threshold for the system is obtained requiring the
largest eigenvalues to be larger the $0$, which leads to the condition
for the presence of an endemic state: 
\begin{equation}
\label{t_m1}
\frac{\lambda}{\mu}> \frac{1}{m}\frac{1}{\langle a
  \rangle+\sqrt{\langle a^2 \rangle}} 
\end{equation}

From this last expression we can
recover the epidemic threshold of Eq.~(\ref{thre}) by considering
$\beta= \lambda \langle k \rangle$, $a_i = \eta x_i$ and $\langle k
\rangle = 2 m \eta \langle x \rangle $.

\section*{Acknowledgments}
  The work has been partly sponsored by the Army Research Laboratory
  and was accomplished under Cooperative Agreement Number
  W911NF-09-2-0053. RPS acknowledges financial support from the
  Spanish MICINN (project FIS2010-21781-C02-01) and additional support through
  ICREA Academia, funded by the Generalitat de Catalunya.
\subsection*{Author Contributions }
 R.PS \& A.V designed research, N.P performed simulations, N.P, B.G, R.PS \& A.V analyzed the data, N.P, R.PS \& A.V contributed new analytical results. All authors wrote,
reviewed and approved the manuscript.
\subsection*{Competing financial interests}
The authors declare no competing financial interests.


\end{document}